\documentclass[iicol]{sn-jnl}

\usepackage{graphicx}%
  \graphicspath{{figures/}}
\usepackage{multirow}%
\usepackage{amsmath,amssymb,amsfonts}%
\usepackage{amsthm}%
\usepackage{mathrsfs}%
\usepackage[title]{appendix}%
\usepackage{xcolor}%
\usepackage{textcomp}%
\usepackage{manyfoot}%
\usepackage{booktabs}%
\usepackage{algorithm}%
\usepackage{algorithmicx}%
\usepackage{algpseudocode}%
\usepackage{listings}%
\usepackage{subcaption}
\usepackage[backend=bibtex,style=numeric]{biblatex}
\begin{document}

\title{Reducing Shape-Graph Complexity with Application to Classification of Retinal Blood Vessels and Neurons}


\author*[1]{\fnm{Benjamin} \sur{Beaudett}}\email{bbeaudett@fsu.edu}

\author[1]{\fnm{Anuj} \sur{Srivastava}}\email{anuj@stat.fsu.edu}


\affil*[1]{\orgdiv{Statistics Department}, \orgname{Florida State University}, \orgaddress{\street{117 N. Woodward Ave}, \city{Tallahassee}, \postcode{32306}, \state{Florida}, \country{USA}}}



\abstract{Shape graphs are complex geometrical structures commonly found in biological and anatomical systems. A shape graph is a collection of nodes, some connected by curvilinear edges with arbitrary shapes. Their high complexity stems from the large number of nodes and edges and the complex shapes of edges. With an eye for statistical analysis, one seeks low-complexity representations that retain as much of the global structures of the original shape graphs as possible. 
This paper develops a framework for reducing graph complexity using hierarchical clustering procedures that replace groups of nodes and edges with their simpler representatives. 
It demonstrates this framework using graphs of retinal blood vessels in two dimensions and neurons in three dimensions. The paper also presents experiments on classifications of shape graphs using progressively reduced levels of graph complexity. The accuracy of disease detection in retinal blood vessels drops quickly when the complexity is reduced, with accuracy loss particularly associated with discarding terminal edges. Accuracy in identifying neural cell types remains stable with complexity reduction.
}




\maketitle

\section{Introduction}
\label{sec:intro}

This paper studies structures of shape graphs with a focus on their statistical analysis and classification. 
A shape graph is a set of curves in space (edges) along with the points where they intersect or terminate (nodes). Shape graphs generalize traditional graphs by 
assigning locations in space to each node and specifying paths in space for each edge 
instead of just binary adjacency values or scalar weights. In other words, one is also interested in the shapes formed by the edge curves when analyzing and comparing shape graphs. 
Shape graphs contain important information in the context of objects such as networks of blood vessels or neurons, branches and roots of plants, and networks of roads or waterways. 
Spatial tree graphs that prohibit cycles are a special case of shape graphs.

In natural systems, the roles and functionalities of network structures are directly reflected in the details of connectivity and spatial relations that shape graphs capture. Thus, tools for statistical analysis of shape graphs provide a novel avenue for characterizing their roles or detecting irregularities.
One case of interest is the retinal blood vessel (RBV) networks of the human eye.
A great deal of work has been done in recent years to create machine learning (ML) algorithms that can extract these networks from retinal fundus images and diagnose ocular diseases 
\cite{DongRetinalOpticDiseaseScreening22,SonDeepLearnRetinalFundus19,ChoiDeepLearningRetinal17,JiangFundusDiseaseDNN,LeeAIDiabeticRetinopathy21,PoplinCardRiskRetinalFundus18}. 
These systems can either assist human diagnosticians or perform independent diagnoses,
thereby reducing the cost of screening for diseases. Several of these systems are already in clinical use, including at least one approved by the Food and Drug Administration \cite{LeeAIDiabeticRetinopathy21}.
A common approach in these ML classifiers is to feed color fundus images directly to an artificial neural network which makes classifications based on pixel intensities.
While some of these produce very accurate results, details about the decision-making are opaque and do not provide insight as to why a classification was made.
Instead of relying on pixel intensities in the fundus images, we focus on the networks of veins and arteries, represented as shape graphs, and study their structural variability. This approach allows us to seek interpretable statistics of geometric details and use these for analysis.

A significant challenge in the analysis of shape graphs comes from their complexity.
Direct comparison of any two graphs requires the registration of nodes across graphs, termed graph matching.
In the context of retinal blood vessels, graph matching can be used to track changes over time \cite{MottaAlignmentRetinalFundus19} or to quantify the dissimilarity between subjects \cite{BalRBV22}.
Current approaches to graph matching have computational complexity that grows as a third-order polynomial in the numbers of nodes and edges \cite{ZhouFactorizedGraphMatching16}, making it difficult to quantify shape differences when the numbers of nodes and edges become even modestly large.
One way to facilitate the matching is by reducing graph complexity, i.e. producing simpler graphs with fewer nodes and edges but which maintain the gross structure of the originals. This relieves the computational burden and causes the matching to focus on large-scale structures without being slowed by finer details.
One can also improve results of matching graphs with differing complexity levels by simplifying the more complex graph to match the other \cite{BalRBV22}.
A related issue is the level of detail in graphs necessary for classification. Do the smaller, peripheral structures carry relevant information, or is it mainly restricted to the core skeletal parts? 
Graph simplification can also be used to investigate this question.

A number of methods for graph simplification have been developed, with a recent survey provided by Liu et al. \cite{liuSafavi2018graphSummarization}. However, very little work has been done that applies to graphs where both nodes and edges have spatial attributes. The only such method we are aware of is that of Basu Bal et al. \cite{BalRBV22}, which clusters nodes and replaces each cluster with a single new node. 
That purely clustering-based approach gave some intuitive results but left visible room for improvement, having a tendency to create bushy collections of terminal edges emanating from a common node.
In this paper we refine that approach, presenting a method for simplifying shape graphs that performs separate hierarchical clusterings of both nodes and edges as well as directly removing small or redundant terminal edges. The clustered nodes and edges are replaced by sparser representations that are based on means of the cluster members. The choice of metrics for clustering is important and dictates the outcomes.

We assess our method through visualizations and statistics-based classification experiments. These use shape graphs in $\mathbb{R}^2$ extracted from retinal fundus photographs and shape graphs in $\mathbb{R}^3$ formed from digital reconstructions of neurons.
We lay out the approach used for shape graph reduction in sections \ref{sec:background}, \ref{sec:dataProcessing}, and \ref{sec:multiRes}, including demonstrations using RBV and neuron graphs in section \ref{sec:multiRes}. In section \ref{sec:otherMethods} we discuss some alternative ideas. We describe and give results from classification experiments in section \ref{sec:classification}.

\section{Mathematical Background} 
\label{sec:background}

\noindent {\bf Shape Graphs}:
We define a shape graph $G$ in $\mathbb{R}^d$ as a set of nodes $V$ represented by points 
$\{v_i \in \mathbb{R}^d \}_{i=1}^n$ 
and a set of edges $B$ represented by parameterized curves 
$\{\beta_{ij}: [0,1]\rightarrow \mathbb{R}^d \}_{i,j=1}^n$ 
with $\beta_{ij}(0)=v_i$ and $\beta_{ij}(1)=v_j$ for adjacent nodes or $\beta_{ij} = 0$ for nonadjacent nodes.
We use the letter $m$ to indicate the number of nontrivial edges in a graph and write $\beta$ with a single index or none at all when the nodes are not being emphasized.
We ensure that all graphs we work with consist of a single connected component. For most of this paper there are no other topological constraints. We refer to the special case where cycles are prohibited as tree graphs.

\noindent {\bf Curve Shape Means}:
To analyze shapes of edge curves, we utilize elements of elastic shape analysis of Euclidean curves. 
We briefly describe the approach here and refer to reader to \cite{JoshiNovRepRiemann07,KurtekSigEstTimeWarp11,SrivastavaShpAnlElCurv11,SrivastavaRiemAnlProbDens07,TuckerProteomics14,TuckerGenModFuncDat13,BalRBV22,GuoBrainArtNet22} for further details.
This approach is based on a mathematical representation called square-root velocity function or \emph{srvf}.
The \emph{srvf} of an edge $\beta$ is defined by an operator 
$
q: \mathcal{C}([0,1],\mathbb{R}^d) \rightarrow \mathbb{L}^2([0,1],\mathbb{R}^d)
$
with 
$$
q_{\beta}(t) := \dot{\beta}(t)/\sqrt{\Vert \dot{\beta}(t) \Vert}.
$$
Here $\mathcal{C}([0,1],\mathbb{R}^d)$ represents the set of all absolutely continuous, $\mathbb{R}^d$-valued functions on $[0,1]$.
Working with the derivative $\dot{\beta}$ immediately removes irrelevant information about location, and using the \emph{srvf} has the additional advantage of providing a notion of shape distance between edges which is invariant to reparameterization. 
Let $\Gamma$ be the group of orientation-preserving diffeomorphisms of $[0,1]$. Reparameterization of an edge by some $\gamma\in\Gamma$ acts on the \emph{srvf} as 
$$
(q_{\beta},\gamma)(t) := q_{\beta\circ\gamma}(t) = q_{\beta}(\gamma(t))\sqrt{\dot{\gamma}(t)}.
$$
A key property of \emph{srvf} representation is the invariance 
$
\Vert (q_{\beta},\gamma) \Vert_2 
= \Vert q_{\beta} \Vert_2 
= \sqrt{\ell},
$
the square root of the arc length of $\beta$, for any $\gamma \in \Gamma$.
Changing the subscript notation to reflect indices, 
transforming two \emph{srvf}s using the same reparameterization yields an unchanged
$
\Vert (q_1,\gamma)-(q_2,\gamma) \Vert_2
= \Vert q_1-q_2 \Vert_2.
$
Next, one defines a quotient space in the \emph{srvf} space according to 
$$
[q]:=\{(q,\gamma)\vert \gamma\in\Gamma \}.
$$
The quotient point $[q_{\beta}]$ represents the set of all curves that are within translations and reparameterizations of $\beta$, and constitutes a shape equivalence class.
The set of all equivalence classes is the shape space ${\cal S} \equiv \mathbb{L}^2([0,1],\mathbb{R}^d)/\Gamma$. This shape space is endowed a metric $d_s$ according to
$$
d_{s}([q_1],[q_2])
:= \operatorname*{min}_{\gamma\in\Gamma} \Vert q_1-(q_2,\gamma) \Vert_2\ .
$$
This $d_{s}$ is a proper distance on ${\cal S}$ and is invariant to both rigid translation and reparameterization of curves \cite{SrivastavaFuncShapeDatAn16}. 
The optimizing $\gamma$ can be approximated using the dynamic programming algorithm.

We can now define a representative shape for a set of \emph{srvf}s $\{q_i \}_{i=1}^I$ as the Karcher or intrinsic mean
$$
[\mu_q] := \operatorname*{argmin}_{[q]} \sum_{i=1}^I d_{s}([q],[q_i])^2.
$$
We find a value $\mu_q$ using a slight alteration of the \emph{srvf} Karcher mean algorithm presented in \cite{SrivastavaRegFuncDatFR11}. After selecting one of the $q_i$ as an initial $\mu_q^1$, the algorithm proceeds iteratively, at each step $k$ reparameterizing the \emph{srvf}s as $\{q_i^k\}_{i=1}^I$ to optimally match $\mu_q^k$ and updating $\mu_q^{k+1} = \frac{1}{I}\sum_{i=1}^I q_i^k$. 
This algorithm is known to converge with monotonically decreasing error, 
but it can converge to a local optimum rather than a true mean.
We use the $\mu_q$ produced by the algorithm without further reparameterization.

Having found a mean \emph{srvf} $\mu_q$ for a set of edges $\{\beta_i \}_{i=1}^I$, an unpositioned mean edge is formed by converting it 
into a curve by integrating 
$
\beta_{\mu}(t) := \int_0^t \mu_q(s)\Vert \mu_q(s) \Vert ds.
$
In our applications we use edge sets $\{\beta_i \}_{i=1}^I$ to form a mean edge 
which always has a prespecified pair of destination nodes $v_a$ and $v_b$ it is intended to connect.
In doing this, we first form a new set of edges $\{\beta_i^* \}_{i=1}^I$ by rotating, translating, and scaling the originals so that they run from $v_a$ to $v_b$. 
These repositioned curves are then used to define $\beta_{\mu}$ which is in turn rotated, translated, and scaled to fit the destination node pair.

\noindent {\bf Chamfer Distance}:
Some steps in our method require measurements of location-based discrepancy between edges, for which the shape distance above cannot be applied. In these cases we use the chamfer distance as defined in \cite{WuChamferDist21}.
In our setting, an edge $\beta_k$ is represented by $T$ consecutive points 
$\{p^{(k)}_i\}_{i=1}^T$.
The chamfer distance between two edges is
\begin{equation}
  d_C(\beta_1,\beta_2) = 
  \frac{1}{T} \sum_{i=1}^T \operatorname*{min}_{p \in \beta_2}\Vert p^{(1)}_i-p\Vert
  + 
  \frac{1}{T} \sum_{i=1}^T \operatorname*{min}_{p \in \beta_1}\Vert p^{(2)}_i-p\Vert
  .
\label{eq:chamferDistance}
\end{equation}
We note that this is not a proper distance since it does not obey the triangle inequality.

\noindent {\bf Effective Resistance}:
Our node clustering method requires a distance between nodes that is attentive to the network structure. We use the effective resistance defined in \cite{EllensEGR11}. Rather than simply finding the shortest edge-based paths between nodes, this distance measures nodes as being progressively closer as the overall connectivity between them is increased.
Take a graph composed of nodes $\{v_i\}_{i=1}^n$ and edges $\{\beta_{ij} \}_{i,j=1}^n$ between them. Let $\ell_{ij}$ be the length of $\beta_{ij}$ and define the weight $w_{ij}=\frac{1}{\ell_{ij}}$ if nodes $v_i,v_j$ are adjacent and $w_{ij}=0$ otherwise. 
Let $\mathbf{W}$ be the $n\times n$ matrix of weights, $\mathbf{S}$ be a diagonal matrix of the strengths $S_{ii} = \sum_{j=1}^n w_{ij}$, and $\mathbf{Q}$ be the weighted Laplacian $\mathbf{Q}=\mathbf{S}-\mathbf{W}$. For a connected graph, this matrix is positive semidefinite with a single zero eigenvalue which has the equiangular vector $\mathbf{1}$ as its eigenvector \cite{KleinResDist93}. The effective resistance between two nodes is then defined as
\begin{equation}
  d_E(v_i,v_j) = (\mathbf{e}_i-\mathbf{e}_j)^T \mathbf{Q}^{-1} (\mathbf{e}_i-\mathbf{e}_j)^T 
\label{eq:effectiveResistance}
\end{equation}
where $\mathbf{e}_i$ is the column vector with $1$ in its $i$th location and zeros elsewhere, and $\mathbf{Q}^{-1}$ is a matrix which operates as an inverse of $\mathbf{Q}$ on span$\{\mathbf{1}\}^{\perp}$ and as the zero map on span$\{\mathbf{1}\}$. This $d_E$ is a proper distance function \cite{KleinResDist93}.

\section{Data Processing}
\label{sec:dataProcessing}
There are two types of datasets used in this paper: retinal blood vessels (RBVs) and neurons.
We convert input data into shape graphs where each edge is discretized as a piecewise linear curve defined with arbitrary granularity. The results presented here used $T=30$ points per edge.

The RBV shape graphs are extracted from retinal fundus photographs.
The photographs are first converted to binary segmentations where each pixel is either black or white to indicate whether a vessel is present in that location. This segmentation was performed prior to our accessing the data.
We then use the Vessel Tech software \cite{WangVeslTch21} to convert the binary segmentations to shape graphs in $\mathbb{R}^2$. 
The nodes of the graph are the points where vessel branches intersect or terminate. The edges connecting the nodes are the stretches of blood vessel between them. 
We do not distinguish between veins and arteries, and the representation does not consider continuity of vessel segments across intersection points. We do not consider vessel thickness.

The neuron shape graphs are obtained directly from \texttt{.swc} encodings of neuron structures in $\mathbb{R}^3$.

Our approach assumes that input graphs consist of a single connected component. This is a reasonable assumption for our data types. However, imperfections in the data gathering and processing can result in disconnected graphs, so we apply a simple method of automatically connecting the graphs:
A connected component $\tilde{G}^i$ of a disconnected graph $\tilde{G}$ is represented as the union of all of the points in its nodes and edges.
The smallest Euclidean distance is found between any point $p \in \tilde{G}^i$ and any point $q \in \tilde{G}^j \subseteq (\tilde{G}^i)^c$ in any other component, such that at least one of $p$ and $q$ is a node. Components $\tilde{G}^i$ and $\tilde{G}^j$ are then joined by creating an edge between nodes at points $p$ and $q$, creating a new node at one of the points if necessary. The shape of the new edge is determined using the Karcher mean of \emph{srvf}s of nearby edges using the method described in section \ref{sec:background}. This is repeated selecting a new $\tilde{G}^i$ from the redefined components until a single connected graph $G$ is produced.

An example of a binary segmented fundus image and the connected graph extracted from it can be found in the top two images in the leftmost column of Fig.~\ref{fig:cyclicReductionAllSteps}.
The leftmost column of Fig.~\ref{fig:cyclicReduction3D3Angles} shows three views of a neuron shape graph with the full complexity of its original encoding.
Our visualizations use human RBV data taken from the ARIA dataset \cite{ariaData,FarnellBldVsl08} and human neurons from the Narkilahti dataset \cite{NarkilahtiNeuroNetIschaemicConditions22} on NeuroMorpho.org \cite{NeuroMorpho1,NeuroMorpho2}.

\section{Multi-Resolution Shape Graph Representation}
\label{sec:multiRes}

In this section we describe a method that reduces graph complexities while maintaining their general shapes.
We use the term `resolution' to refer to the complexity. Thus, representing a graph at multiple resolutions by progressively decreasing its complexity leads to a `multi-resolution' representation.
The process runs as a cycle consisting of two kinds of steps:
(1) Terminal edge removal: Terminal edges identified as very short or as similar to nearby terminal edges do not contribute much to fundamental structure and are removed.
This helps prevent formation of bushy groups of terminal edges that was seen in earlier work.
(2) Local summarization via clustering: Separate hierarchical clustering steps for nodes and edges identify clusters throughout the graph and replace them with simpler structures.
It is common in our data to find long, unbroken edge segments running alongside each other which are redundant in terms of the skeletal shape of the graph. Including edge-based clustering allows the algorithm to identify and merge these in cases where node clustering alone would leave them unchanged.

All of the trimming and clustering steps, as well as the graph connection in section~\ref{sec:dataProcessing}, can produce degree-two nodes. 
These can often be removed without affecting the graph's spatial structure by simply deleting the node and concatenating the two adjacent edges. 
In some cases though, the two adjacent nodes are themselves adjacent, forming a `triangle' such that removing the degree-two node would leave the others directly connected by two separate edges. 
We remove the former type as they appear throughout the process, but leave the latter unchanged. 
For conciseness, we usually omit mention of these removals in the details below.

All of our hierarchical clustering is performed using the native \texttt{linkage} and \texttt{cluster} functions in MATLAB \cite{MtLb}.

\subsection{Terminal Edge Removal}
\label{ssec:terminalEdgeRemoval}

\begin{figure*}
  \centering
  \includegraphics[width=1\textwidth, scale=1]{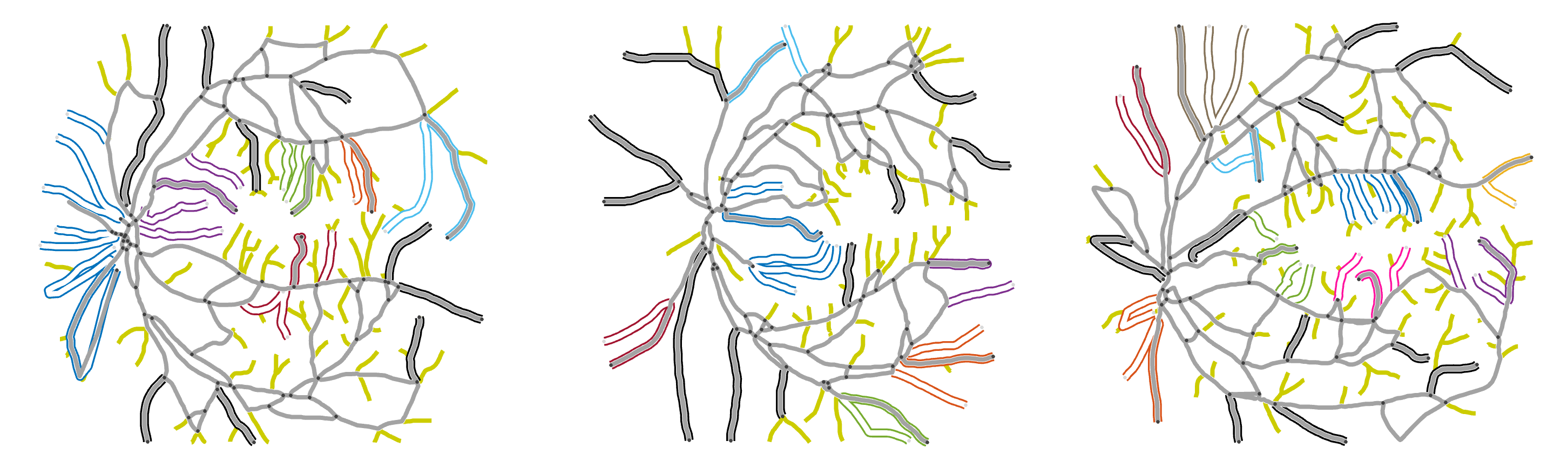}
    \caption{
      Terminal edge removal applied to three RBV networks. 
      Yellow edges are removed during short-terminal trimming. 
      This is followed by similar-terminal trimming. 
      Colored highlights indicate similar-terminal clusters. 
      Black highlights indicate terminal edges not assigned to any cluster.
      Only the longest edge in each cluster is retained, with the white edges being removed.
      Gray edges show the graph that remains after trimming.
    }
    \label{fig:terminalTrimming}
  \includegraphics[width=1\textwidth, scale=1]{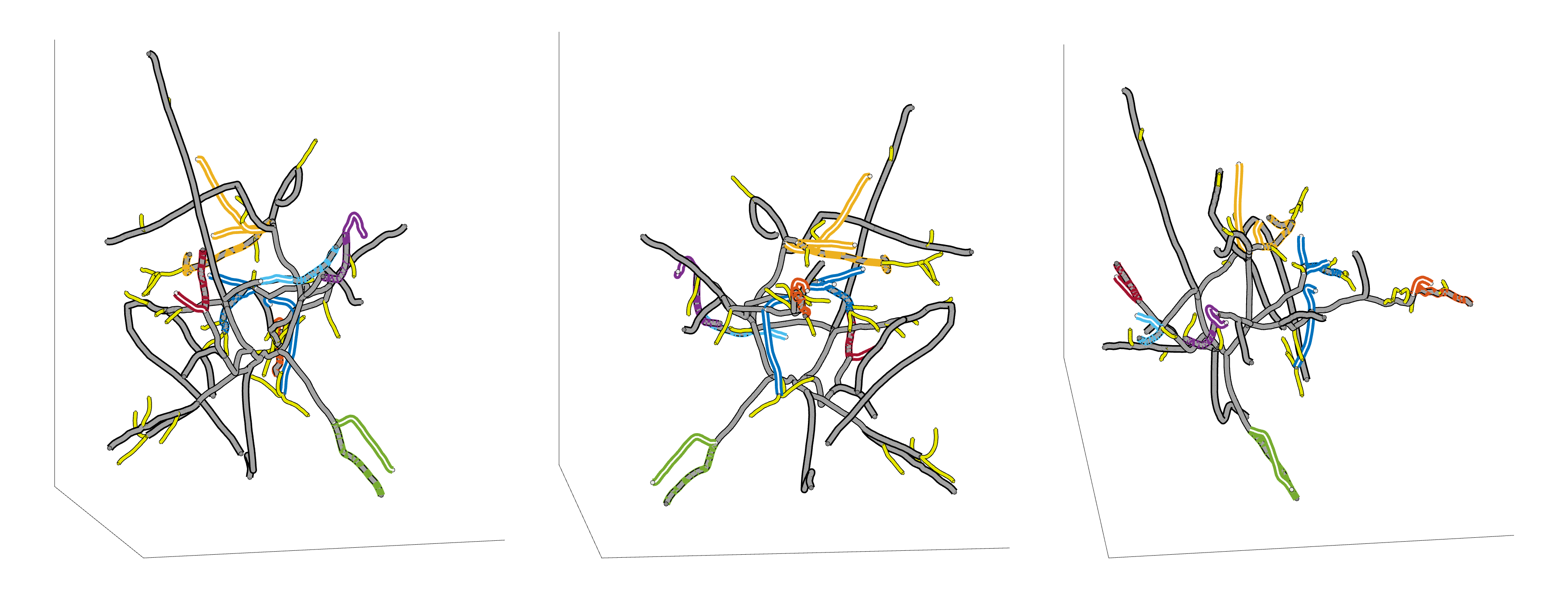}
    \caption{
      Terminal edge removal applied to a neuron. Three views of the same graph.
      Bright yellow edges are removed during short-terminal trimming. 
      This is followed by similar-terminal trimming. 
      Colored highlights indicate similar-terminal clusters. 
      Thicker black highlights indicate terminal edges not assigned to any cluster.
      Only the longest edge in each cluster is retained, with the white edges being removed.
      Gray edges show the graph that remains after trimming.
    }
    \label{fig:terminalTrimming3D}
\end{figure*}

We consider two types of terminal edges that can be removed with only minor changes in graph structure: short terminal edges and groups of terminal edges that are similar to each other in terms of shape and location. We perform two separate processes to remove these kinds of edges from a graph. 
The first uses information about the graph to define a length threshold for a short edge then removes terminal edges that are shorter. The second uses clustering to identify groups of similar terminal edges then removes all but the longest from each group.
We outline these processes in algorithms \ref{alg:shortTerminalTrimming} and \ref{alg:similarTerminalTrimming}.
For the similar terminal removal we use the chamfer distance $d_C$ from equation \ref{eq:chamferDistance} and single-linkage (shortest distance) hierarchical clustering.
Whenever a round of terminal trimming is applied, it includes short terminal removal followed by similar terminal removal.
Application of the terminal removal steps is illustrated in Figs. \ref{fig:terminalTrimming} and \ref{fig:terminalTrimming3D}.

\begin{algorithm} 
\caption{
  Short terminal edge trimming
}
  \begin{algorithmic}[1] 
    \State \textbf{Input:}
      A shape graph with $n$ edges, and tuning parameters $\theta_{tag}$ and $\theta_{til}$.
    \State Compute the lengths of all $n$ edges in the graph.
    \State Set $\ell_{\theta_{til}}$ as the $\theta_{til}$ percentile of the lengths.
    \State Set a length threshold $\ell_{\theta}=\theta_{tag} * \ell_{\theta_{til}}$ as a percentage of the percentile.
    \State \textbf{while} the graph contains terminal edges with length less than $\ell_{\theta}$:
    \State Remove terminal edges with length less than $\ell_{\theta}$ along with their terminal nodes.
        Removing terminal edges redefines which remaining edges are terminal, so this must be done repeatedly until the condition is met.
        Also, removing a terminal edge often reduces the degree of a node from three to two. In order to avoid removing chains of short terminal edges which are more properly considered a single long one, degree-two nodes are detected and elided as appropriate between rounds of edge removal.
    \State \textbf{end while}
    \State \textbf{Output:}
      A shape graph with short terminal edges and associated nodes removed, and all other nodes and edges identical to the input.
  \end{algorithmic}
  \label{alg:shortTerminalTrimming}
\end{algorithm}

\begin{algorithm} 
\caption{
  Similar terminal edge trimming
}
  \begin{algorithmic}[1] 
    \State \textbf{Input:}
      A shape graph with $m$ edges, a tuning parameter $\phi_{til}$, a distance function defined on shape graph edges, and a distance-based clustering method that determines the number of clusters based on the data and can restrict cluster formation according to a distance ceiling.
    \State Compute the distances between all $m$ edges in the graph.
    \State Set a maximum cluster radius $rad_{\phi}$ equal to the $\phi_{til}$ percentile of the distances.
    \State Perform clustering on only the terminal edges. Clustering is performed to the extent possible without allowing cluster formation at distances above $rad_{\phi}$.
    \State Identify the longest edge in each cluster (including singletons) and remove all other terminal edges along with their terminal nodes.
    \State \textbf{Output:}
      A shape graph with visually redundant terminal edges and associated nodes removed, and all other nodes and edges identical to the input.
  \end{algorithmic}
  \label{alg:similarTerminalTrimming}
\end{algorithm}

\subsection{Node Clustering}
\label{ssec:nodeClust}

\begin{figure*}[ht]
  \centering
  \includegraphics[width=1\textwidth, scale=1]{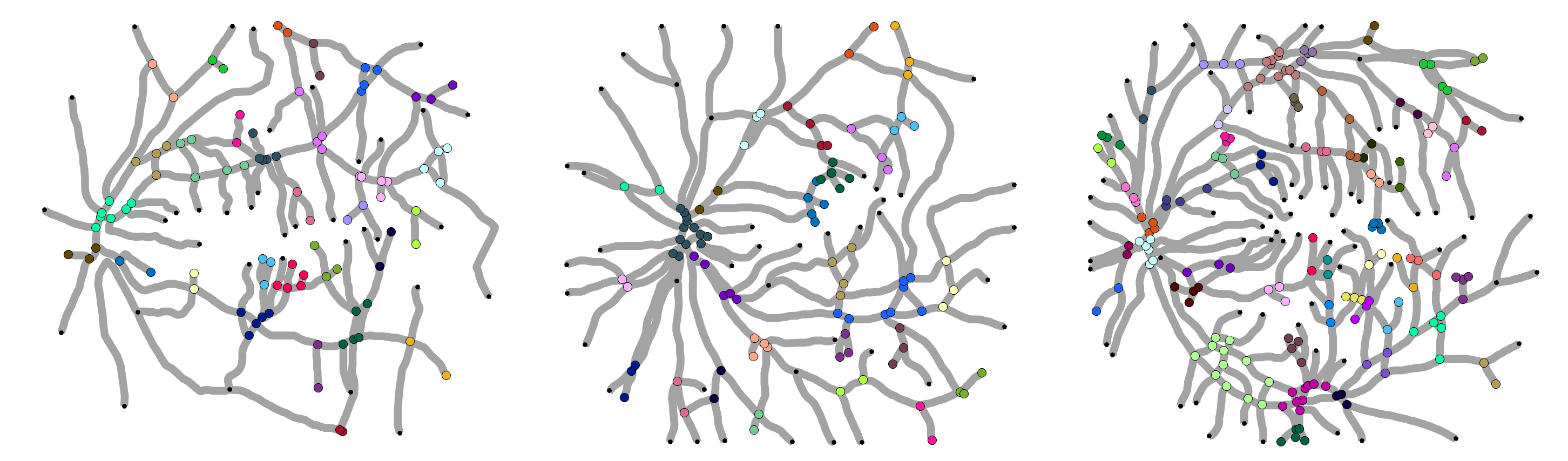}
    \caption{
      Node clusters identified in three RBV networks using a .5 resolution target $n_r=0.5n$.
      Nodes marked with the same color form a cluster.
      Smaller black nodes are not assigned to any cluster.
      The reduced graphs are not shown in this figure.
    }
    \label{fig:clusteringHighlighting_nodes}
  \includegraphics[width=1\textwidth, scale=1]{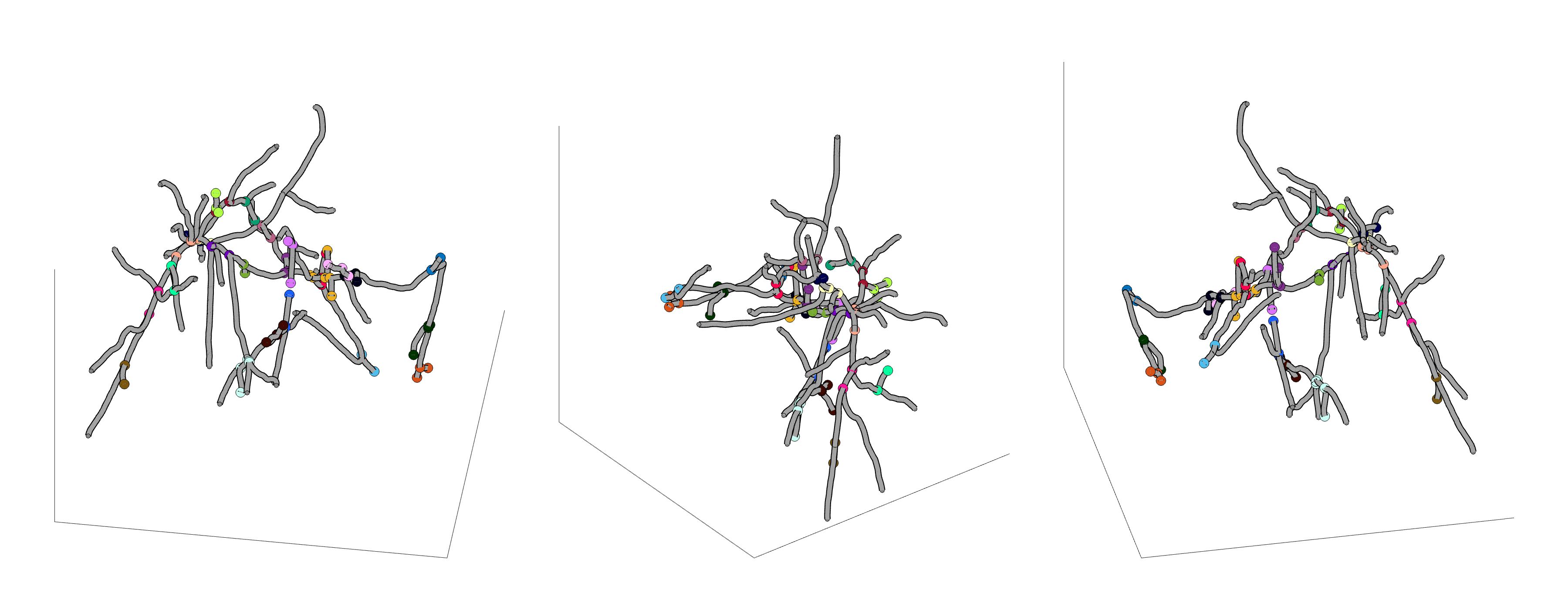}
    \caption{
      Node clusters identified in a neuron using a .5 resolution target $n_r=0.5n$. Three views of the same graph.
      Nodes marked with the same color form a cluster.
      Smaller gray nodes are not assigned to any cluster.
      The reduced graph is not shown in this figure.
    }
    \label{fig:clusteringHighlighting_nodes3D}
\end{figure*}

The node-based clustering performed here is the same as used in \cite{BalRBV22}. 
We outline the process in algorithm \ref{alg:nodeClustering}.
For distance we use the effective resistance $d_E$ from equation \ref{eq:effectiveResistance}, and our distance-based clustering method is complete-linkage (largest distance) hierarchical clustering.
Figs.~\ref{fig:clusteringHighlighting_nodes} and \ref{fig:clusteringHighlighting_nodes3D} illustrate the node clustering produced by this method.

\begin{algorithm}
\caption{
  Node-clustering reduction of a shape graph
}
  \begin{algorithmic}[1] 
    \State \textbf{Input:}
      A shape graph with $n$ nodes, a target reduced node count $n_r<n$, a distance function defined on shape graph nodes, and a choice of distance-based clustering method that allows the user to specify the number of clusters produced.
    \State Compute the distances between all $n$ nodes in the graph.
    \State Perform clustering on the nodes such that the number of multiple-node clusters plus the number of singletons is equal to $n_r$. These two types are treated equivalently as clusters.
    \State For each cluster $c_i$, form a new node $v_i^*$ located at the Euclidean mean position of the nodes in the cluster.
    \State Determine adjacencies between the new nodes. For each cluster $c_i,$ the new node $v_i^*$ representing it is made adjacent to all nodes $v_j^*$ representing clusters $c_j$ for which there was any prior adjacency between some nodes $v_i\in c_i$ and $v_j\in c_j$.
    \State Form edges between adjacent new nodes $v_i^*$ and $v_j^*$ using the \emph{srvf} Karcher mean described in section~\ref{sec:background}. The mean shape is defined using all edges connecting a node in $c_i$ to a node in $c_j$, and $v_i^*$ and $v_j^*$ are the destination nodes. 
    \State \textbf{Output:}
      A shape graph defined by the new nodes $\{v_i^*\}_{i=1}^{n_r}$ and the edges connecting them.
  \end{algorithmic}
  \label{alg:nodeClustering}
\end{algorithm}

\subsection{Edge Clustering}
\label{ssec:edgeClustering}

\begin{figure*}
  \centering
  \includegraphics[width=1\textwidth, scale=1]{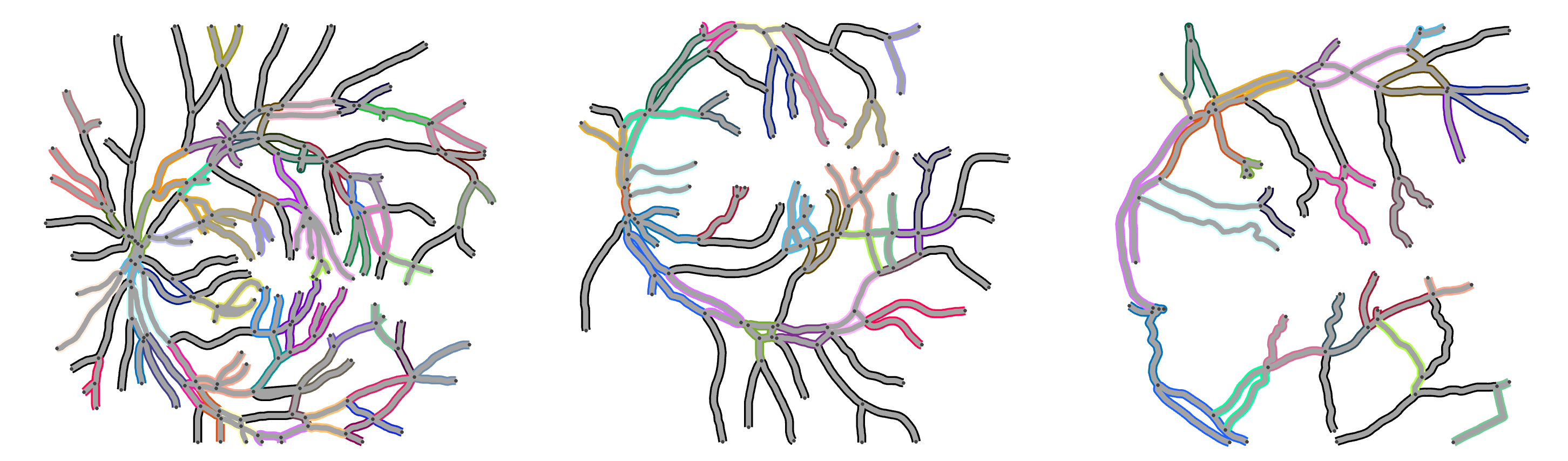}
    \caption{
      Edge clusters identified in three RBV networks using a .4 resolution target $m_r=0.4m$.
      Edges marked with the same color form a cluster.
      Black highlights indicate edges not assigned to any cluster. 
      The reduced graphs are not shown in this figure.
    }
    \label{fig:clusteringHighlighting_edges}
  \includegraphics[width=1\textwidth, scale=1]{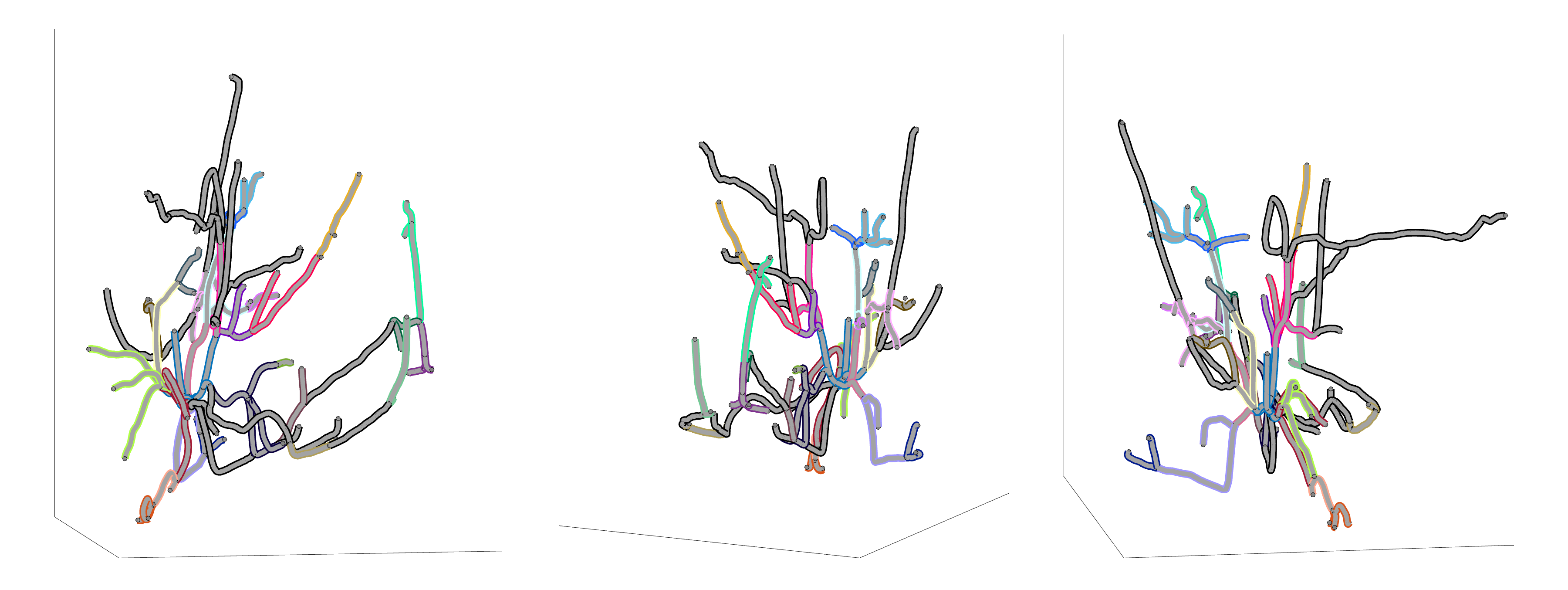}
    \caption{
      Edge clusters identified in a neuron using a .4 resolution target $m_r=0.4m$. Three views of the same graph.
      Edges marked with the same color form a cluster.
      Black highlights indicate edges not assigned to any cluster. 
      The reduced graph is not shown in this figure.
    }
    \label{fig:clusteringHighlighting_edges3D}
  \centering
  \includegraphics[width=1\textwidth, scale=1]{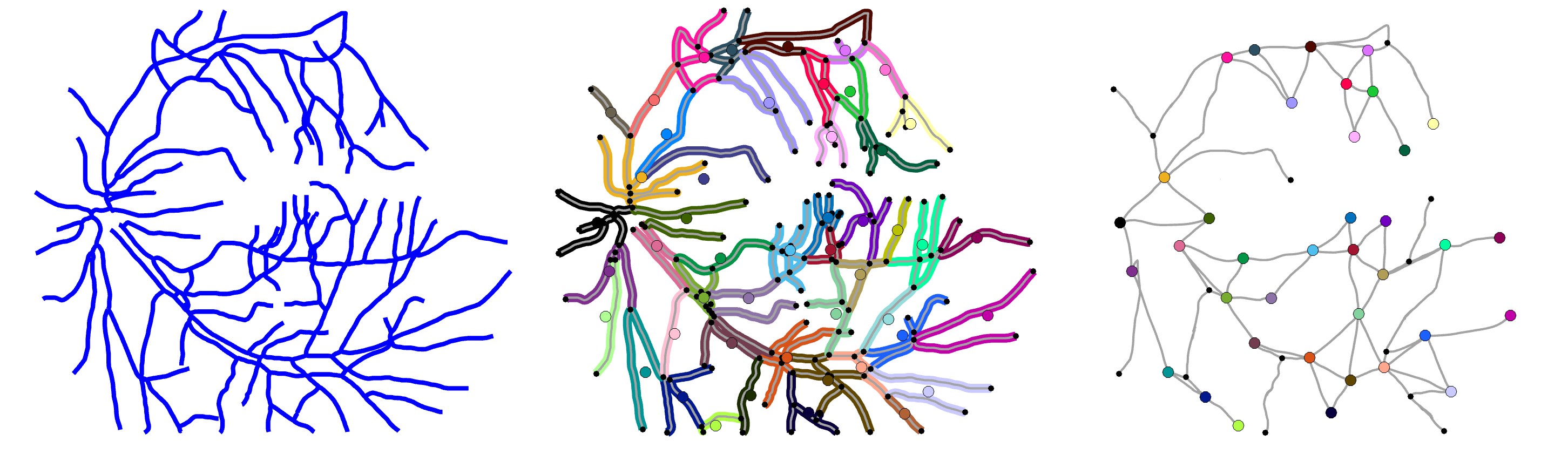}
    \caption{     
      Illustration of the edge clustering reduction step applied to a single RBV network.
      \emph{Left}: The unreduced graph.
      \emph{Center}: Edge clustering of the unreduced graph. Colored highlights indicate identified clusters. Colored dots indicate the node mean for the corresponding cluster.
      \emph{Right}: The reduced graph produced by the edge clustering step, used as an intermediate step in the reduction cycle. Colored dots are inherited from the center plot and indicate nodes that have been created to represent edge clusters. 
      Smaller black dots are nodes retained from singleton edges.
    }
    \label{fig:edgeClusteringReduction}
\end{figure*}

The method used for edge-based clustering is similar to that used for the nodes, with the distinction that an edge cluster is not directly replaced by an edge but rather is replaced by a node. Thus note that while the clustering uses a target number $m_r$ of edge clusters, this value is more directly associated with the number of nodes in the output graph than the number of edges.
We outline the process in algorithm \ref{alg:edgeClustering}.
For distance we use the chamfer distance $d_C$ from equation \ref{eq:chamferDistance}, and our distance-based clustering method is average-linkage (arithmetic mean) hierarchical clustering.
Figs.~\ref{fig:clusteringHighlighting_edges}, \ref{fig:clusteringHighlighting_edges3D}, and \ref{fig:edgeClusteringReduction} illustrate the edge clustering produced by this method.

\subsection{The Multi-resolution Reduction Cycle}
\label{ssec:reductionCycle}

Through exploratory experiments and visual assessment of results of the resolution reduction methods described above, we noticed two trends:
First, applying resolution reduction as a sequence of terminal edge trimming, edge clustering, node clustering, and a further terminal trimming in that order was effective in producing skeletal reductions at the target complexity.
Second, when the degree of complexity reduction was large, the skeletal structure was better maintained if the reduction was performed as a sequence of gradual reductions as opposed to fully dropping the resolution in a single step.
In regard to the first observation, we note that while the node clustering step is guaranteed to produce a reduced graph with no more than the number $n_r$ of target nodes, the edge clustering reduction generally produces graphs with somewhat more nodes than the number $m_r$ of target clusters, and can possibly even produce graphs larger than the input. While this would apparently be a flaw if using edge clustering as a standalone reduction method, it is a useful feature when used in conjunction with the node clustering. The edge clustering finds collections of similar edges and replaces them with highly-interconnected nodes which are then condensed by the node clustering, which enforces the desired node count in the reduced graph. We found that using edge clustering followed by node clustering gave more visually convincing results than using node clustering alone.

\begin{figure*}
    \centering
    \includegraphics[width=1.0\textwidth, scale=1]{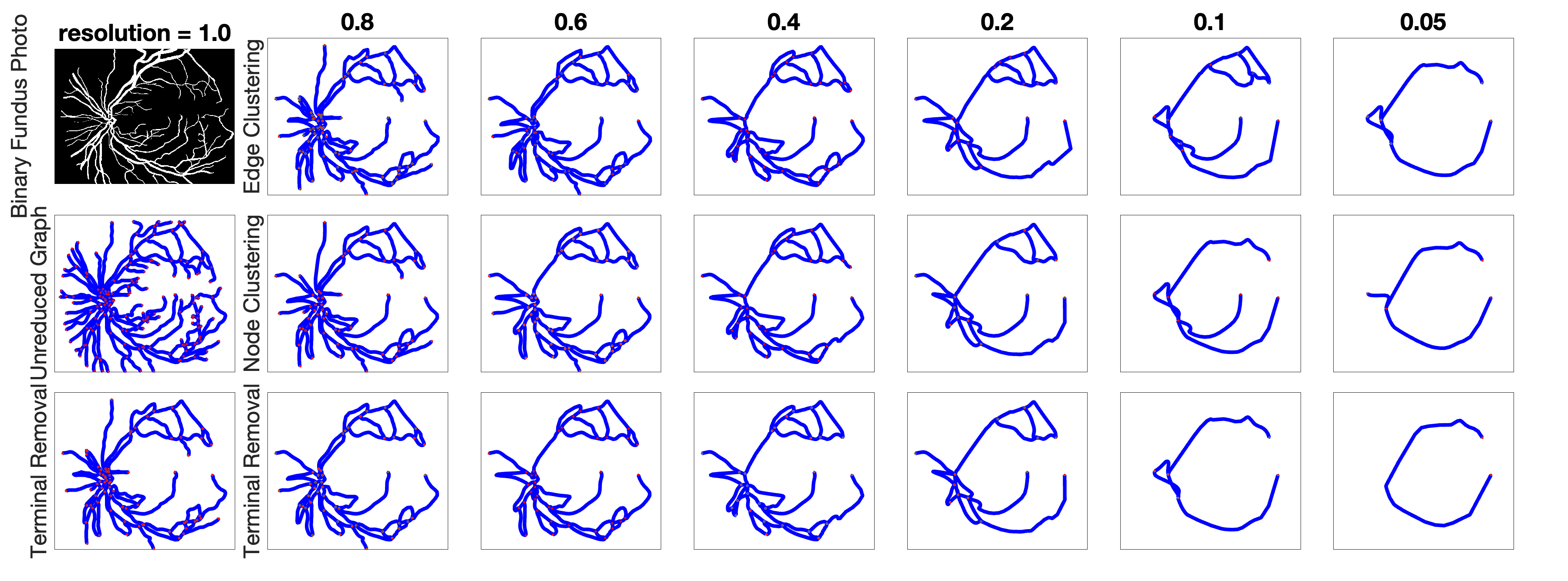}
    \caption{
    The multi-resolution reduction process applied to one RBV network using resolutions $\rho_r\in\{1,.8,.6,.4,.2,.1,.05\}$, including intermediate steps.
    Columns progress from left to right and rows within each column progress from top to bottom.
    \emph{First column:}
      Binary segmented retinal fundus photograph (top), 
      connected graph $G$ extracted from it (middle), 
      and $G_0$ produced by preliminary trimming of terminal edges (bottom).
    \emph{Columns with resolution $\rho_r<1$:}
      Graphs 
        $\tilde{G}_{r}^{edge}$ (top), 
        $\tilde{G}_{r}^{node}$ (middle), 
        and $G_{r}$ (bottom) 
        produced by sequential edge clustering, node clustering, and terminal trimming.
    }
    \label{fig:cyclicReductionAllSteps}
\end{figure*}

\begin{figure*}
    \centering
    \includegraphics[width=1.0\textwidth, scale=1]{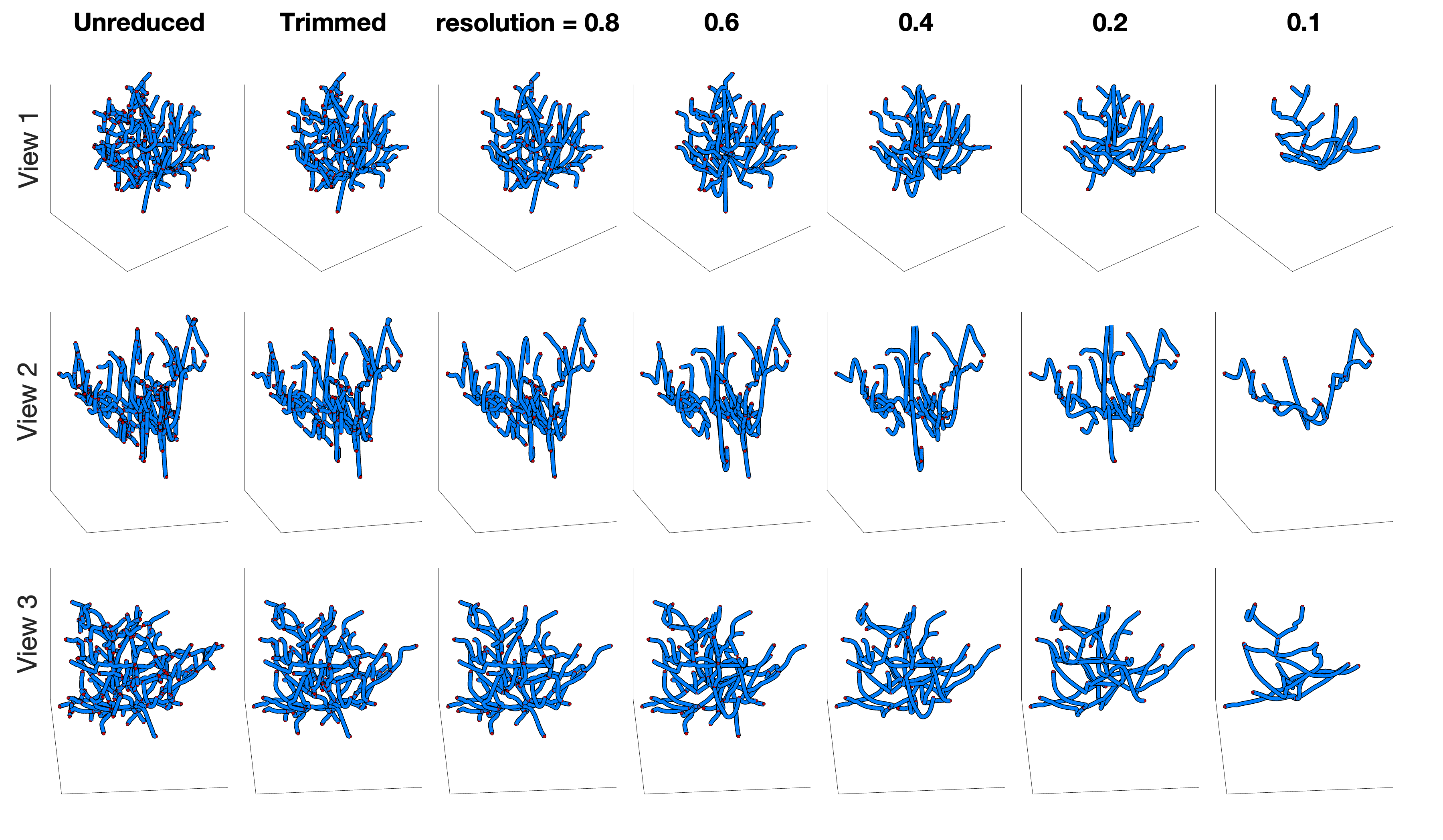}
    \caption{
    The multi-resolution reduction process applied to a neuron using resolutions $\rho_r\in\{1,.8,.6,.4,.2,.1\}$. Each column shows three views of the same graph $G_r$ for the indicated resolution level.
    The intermediate substeps of the reductions are not displayed in this figure.
    }
    \label{fig:cyclicReduction3D3Angles}
\end{figure*}

We perform complexity reduction using a multi-resolution sequence. This consists of performing edge clustering, node clustering, and terminal edge trimming cyclically to produce reduced graphs at progressively decreasing resolution levels $\rho_0 = 1 > \rho_r > \rho_{r+1} > 0$.
The complete process of producing a multi-resolution reduction of a shape graph using a set of resolutions $\{\rho_r\}_{r=0}^R$ is detailed below, summarized in algorithm \ref{alg:multiResolution}, and illustrated in Figs.~\ref{fig:cyclicReductionAllSteps} and \ref{fig:cyclicReduction3D3Angles}.

The input shape graph $G$ first undergoes a preliminary round of terminal edge removal, producing a graph $G_0$.
The numbers $m_0$ and $n_0$ of nontrivial edges and nodes in $G_0$ are used to determine edge cluster targets $\{m_r = \lceil \rho_r * m_0 \rceil\}_{r=1}^R$ and node count targets $\{n_r = \lceil \rho_r * n_0 \rceil\}_{r=1}^R$ for the clustering steps at each resolution level.
Then the resolution reduction sequence is performed for each resolution $\rho_r,\,r\ge 1$.
  First, edge clustering is performed on $G_{r-1}$ with an edge cluster target of $m_r$ to produce $\tilde{G}_{r}^{edge}$.
  Next, node clustering is performed on $\tilde{G}_{r}^{edge}$ with a node count target of $n_r$ to produce $\tilde{G}_{r}^{node}$. 
  The sequence is completed by applying terminal edge trimming to $\tilde{G}_{r}^{node}$ to produce $G_{r}$, the reduced-complexity graph for that resolution.

The graph produced at a given reduced resolution is dependent on the sequence of resolutions included in the process. For example, a graph produced by reducing directly from resolution 1.0 to 0.6 will generally differ from the one produced by reducing the resolution progressively from 1.0 to 0.8 to 0.6. 
For our data, we found that progressive reduction using resolution step sizes of 0.2 usually produced satisfactory results for reduction down to resolution 0.4 or lower. 

\begin{algorithm} 
\caption{
  Edge-clustering reduction of a shape graph
}
  \begin{algorithmic}[1] 
    \State \textbf{Input:}
      A shape graph with $m$ edges, a target number of edge clusters $m_r<m$, a distance function defined on shape graph edges, and a choice of distance-based clustering method that allows the user to specify the number of clusters produced.
    \State Compute the distances between all $m$ edges in the graph.
    \State Perform clustering on all edges in the graph such that the number of true multiple-edge clusters plus the number of singletons is equal to $m_r$. The singletons are treated differently from the true clusters in what follows.
    \State For each singleton edge, retain its two endpoints as nodes $v_i^*$ connected by the original edge without change.
    \State For each true edge cluster $c_i$, form a new node $v_i^*$ located at the Euclidean mean position of all unique nodes that served as an endpoint for any of the edges in the cluster.
    \State Determine adjacencies involving newly-defined nodes. 
      For new nodes $v_i^*$ and $v_j^*$ both of which are associated with true clusters, if there is any node $v_{ij}$ in the input graph which serves as an endpoint for an edge in $c_i$ and an edge in $c_j$, the new nodes are made adjacent.
      If $v_i^*$ is retained from a singleton and $v_j^*$ is formed from a true cluster, they are made adjacent if there is any edge connected to $v_i$ which is a member of $c_j$.
    \State Form edges associated with newly-defined nodes.
      Each new edge is an \emph{srvf} Karcher mean as described in section~\ref{sec:background}.
      When both $v_i^*$ and $v_j^*$ are defined from true clusters, the mean edge shape is defined using all edges in $c_i$ and $c_j$ connected to any node $v_{ij}$ which serves as an endpoint to at least one edge in each cluster.
      If $v_i^*$ is retained from a singleton and $v_j^*$ is formed from a true cluster, the mean edge shape is defined using the singleton edge and all edges in $c_j$ which have $v_i$ as an endpoint.
      In both cases, $v_i^*$ and $v_j^*$ are the destination nodes.
    \State \textbf{Output:}
      A shape graph defined by the new and retained nodes $\{v_i^*\}_{i=1}^{\tilde{m}_r}$ and the edges connecting them, where $\tilde{m}_r$ is equal to the number of true edge clusters plus the number of nodes retained from singleton edges.
  \end{algorithmic}
  \label{alg:edgeClustering}
\end{algorithm}

\begin{algorithm}
\caption{
  Multi-resolution reduction of a shape graph 
}
  \begin{algorithmic}[1] 
    \State
      \textbf{Input:} Connected graph $G$, resolutions $\{\rho_r\}_{r=1}^R$, and trimming parameters 
        $\theta_{tag},\theta_{til}$, and $\phi_{til}$.
    \State
      Trim $G$ to produce $G_0$ and collect its edge count $m_0$ and node count $n_0$.
    \State
      Define edge count targets $\{m_r = \lceil \rho_r * m_0 \rceil\}_{r=1}^R$ and node count targets $\{n_r = \lceil \rho_r * n_0 \rceil\}_{r=1}^R$ for clustering.
    \State \textbf{for} $r$ from $1$ to $R$:
      \State 
        Perform edge clustering on $G_{r-1}$ to produce $\tilde{G}_{r}^{edge}$.
      \State 
        Perform node clustering on $\tilde{G}_{r}^{edge}$ to produce $\tilde{G}_{r}^{node}$.
      \State
        Perform terminal edge trimming on $\tilde{G}_{r}^{node}$ to produce $G_{r}$.
    \State \textbf{end for}
    \State
      \textbf{Output:} Multi-resolution graph sequence $\{G_r\}_{r=0}^{R}$
  \end{algorithmic}
  \label{alg:multiResolution}
\end{algorithm}

\section{Other Reduction Methods}
\label{sec:otherMethods}

There is no general best approach for simplifying a graph. This is especially true for shape graphs with curve-valued edge attributes. 
We investigated a variety of methods for reducing complexity, with the one exhibited in most of this paper being the one we preferred. 
In this section we describe some others we tried.

\subsection{$k$-means Clustering}
\label{ssec:kMeans}

Rather than using the greedy approach of hierarchical clustering, one might like to have a method that is guided by a global optimization objective.
\footnote{
  Although methods of hierarchical clustering defined by optimality conditions have been devised, most work to date on the topic has focused on applying algorithms without attention to mathematically formalized objectives \cite{CohenKanadeMallmannMathieuHierarchicalClustering2019}.
}
A family of such methods is offered by $k$-means clustering. 
Given some dissimilarity measure on point sets, our objective is to produce a graph with $k$ nodes such that the sum of the dissimilarities between the edges in the input graph and their corresponding nodes in the reduced graph is minimized. 
The reduction is performed as follows: 
To begin, $k$ points in $\mathbb{R}^d$ within a region around the input graph are randomly chosen as initial cluster centers. 
Viewing the discretized edges as point clouds, dissimilarities are computed between each edge and each center. Each edge is assigned to the least dissimilar center. A new center for each of the resulting edge clusters is defined by finding the single point in the nearby region which produces the minimum sum of dissimilarities within its cluster. 
The reassign and re-center steps are repeated until the dissimilarity sum converges to a local minimum.
The final cluster centers are used as a node set for a reduced graph, and the edge clusters are used to construct a new edge set as described in section \ref{ssec:edgeClustering}.

We tried this approach with various dissimilarity methods including chamfer distance, maximum set distance, minimum set distance, and Euclidean distance between centroids. All of these were tested with and without using edge lengths as weights for the dissimilarities. 
We judged the results to be uniformly poor by visual assessment. Having edges cluster based on affinity to a common point rather than to each other creates groupings with some symmetry about that point. This means that edges on opposite sides of the cluster center, despite having only their endpoints close to each other, are frequently grouped together. This often resulted in clusters including edges from distinct branches of the graph, joining together sections that should have remained separate.

\subsection{Pairwise Clustering}
\label{ssec:pairwiseClustering}

The results from $k$-means clustering highlighted the importance of direct edge-to-edge comparisons.
With this in mind, we turned back to the cyclic process from section \ref{sec:multiRes} but trying alternative methods at the edge clustering step.
This included a number of variations of the hierarchical clustering in section \ref{ssec:edgeClustering} as well as a probabilistic pairwise clustering method detailed in \cite{SrivastavaJoshiMioLiuShapeAnalysisClustering2005}. We outline the probabilistic method here.

The algorithm takes a dissimilarity matrix as input, and after performing a random initial clustering it proceeds by performing two kinds of steps. 
The first randomly selects one of the input edges and randomly assigns it to one of the existing clusters. The assignment is based on to a probability vector with weights for each cluster corresponding to what the sums of all pairwise dissimilarities within all clusters would be if the edge were assigned to that cluster. 
Assignments with lower dissimilarity sums are given higher weights.
The second step randomly selects two edges from different clusters and randomly either swaps their cluster assignments or does nothing. 
The decision to swap is based on a two-element probability vector with weights determined by what the sum of all pairwise dissimilarity sums within all clusters would be if the swap were or were not performed.
Performing both of these steps in sequence constitutes one round of the algorithm. 
In each of the two steps, the probability values are modulated by a temperature parameter that makes the probability vectors more uniform when the temperature is high and accentuates the higher probability values when the temperature is low. The temperature is decreased at each round of the algorithm, introducing an element of simulated annealing.
The algorithm runs until completing a prespecified number of rounds (we used 12,000).

Given a clustering of a graph's edges, we again took the approach of representing each edge cluster by a single node and connecting these nodes as in section \ref{ssec:edgeClustering}. This still leaves the question of how to define these new nodes. In addition to using Euclidean means of preexisting nodes as before, we also tried a number of dissimilarity-based cluster centers as with the $k$-means approach, except here the centers were only computed after the clusters were already defined.

We tried both the hierarchical clustering and the probabilistic clustering using a variety of settings. They were run using dissimilarity measures of chamfer distance, maximum set difference, and Euclidean distance of centroids. Each of these were tried with new nodes defined using Euclidean node means or edge dissimilarity centers. A number of these yielded visually satisfactory results. The hierarchical method with chamfer dissimilarity and Euclidean nodes that is the focus of this paper was among the best.

\subsection{Edge Removal}
\label{ssec:edgeRemoval}

A simple approach to reducing graph complexity would be to use a reductive process which works entirely by removing nodes and edges without ever creating new ones. 
This has the advantage that no unwanted features such as novel cycles can be introduced. 
The terminal edge trimming in section \ref{ssec:terminalEdgeRemoval} meets this description and could be effective as a standalone method for simplifying tree graphs. 
However, using a method comprised exclusively of removals to simplify a graph where cycles are allowed presents some hurdles.
In our data it was common to see cycles or series of cycles composed of edge segments running parallel to each other and joined by other edges bridging across them
(see Figs.~\ref{fig:terminalTrimming},\ref{fig:clusteringHighlighting_nodes}, and \ref{fig:clusteringHighlighting_edges} for reference). 
In these cases an ideal simplification would remove the cycles and represent the entire region with a single edge. Clearly, then, a removal algorithm must be allowed to consider non-terminal edges. 
Safeguards would need to be in place to prevent the graph from becoming disconnected.
This would not be too difficult, but there is the more nuanced problem of how to select removals that leave a sensible graph behind. 
Removing edges from these cycles using length-based criteria would be prone to producing jagged zig-zag paths that jump back and forth across both sides of a gap, giving a poor representation of the underlying trend.
A more sophisticated approach would be needed to deal with this kind of issue. 
We did not explore this direction any further, but feel it would be worthwhile to pursue.

\subsection{Medial Axis}
\label{ssec:medialAxis}

We also experimented with simplification using medial axes \cite{LieutierWintraeckenGromovHausdorffMedialAxis2023} of sets defined by a shape graph.
The basic idea is to broaden the graph into a wider region and use the medial axis of the border of that region as a reduced graph.
We begin by performing a distance transform of space relative to a graph $G$ using the function 
$D_G: \mathbb{R}^d\rightarrow \mathbb{R}_{\ge 0}$ with $D_G(x) = \operatorname*{min}_{p\in G} d(x,p)$ 
where $d(\cdot ,\cdot )$ is the Euclidean distance. 
This is used to define level sets $K_r = \{x\in\mathbb{R}^d \vert D_G(x)=r\}$. 
The medial axis $\tilde{M}_r$ of $K_r$ is defined by performing another distance transform
$c_r(x) = \operatorname*{min}_{q\in K_r}d(x,q)$ 
and the set-valued function
$\theta_r(x) = \{q\in K_r \vert d(x,q)=c_r(x)\}$,
then setting
$\tilde{M}_r = \{x\in\mathbb{R}^d \vert \# \theta_r(x) \ge 2\}$. 
This set can contain both an interior subset where $D_G(x)<r$ and an exterior subset where $D_G(x)>r$.
We are interested only in the interior portion, which we call $M_r$.
A multi-resolution reduction of a shape graph can be produced by setting $G_0=G$ and progressively simplifying $G_r=M_r$ as $r$ increases.

\begin{figure*}
  \centering
  \includegraphics[width=1\textwidth, scale=1]{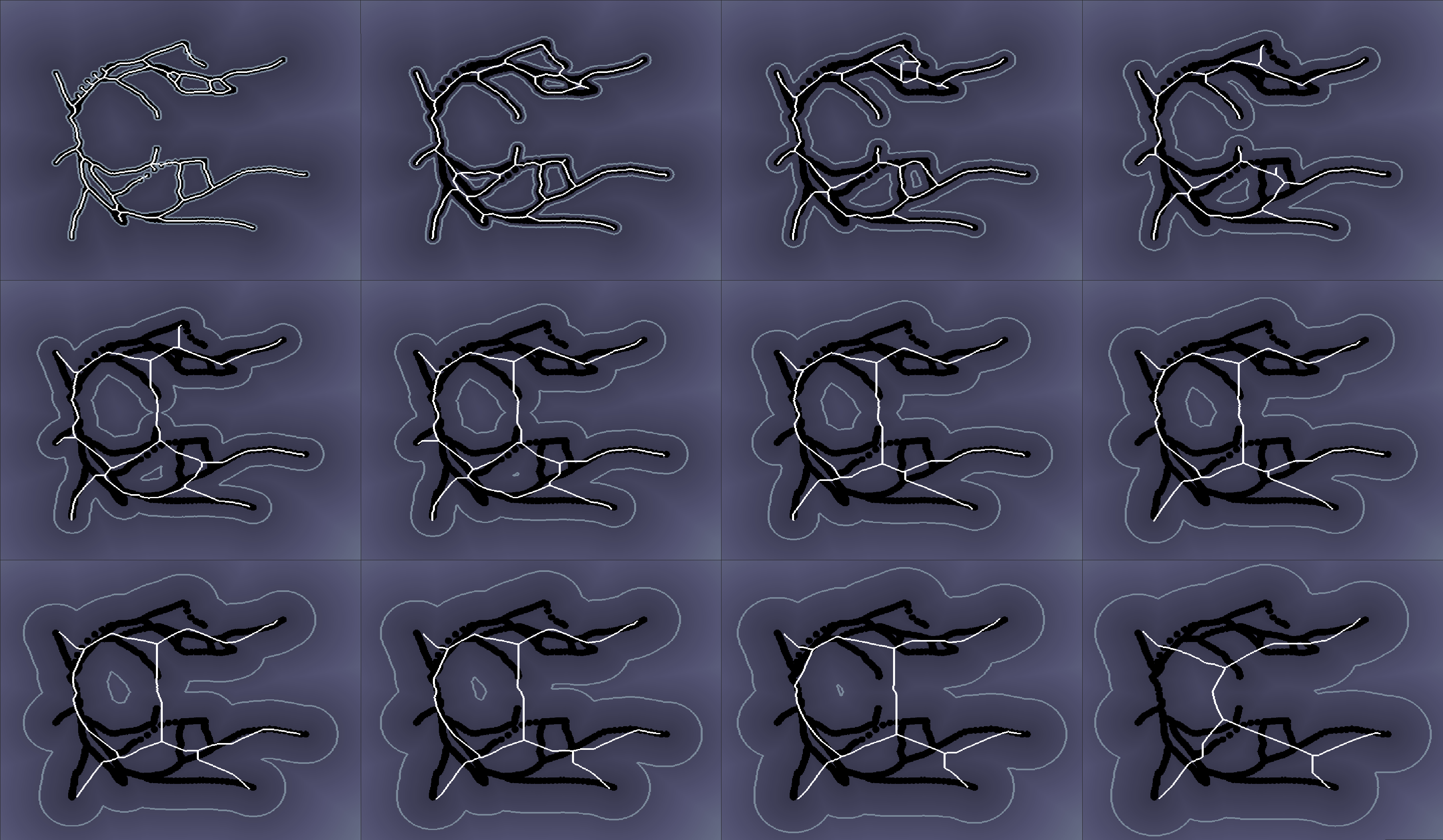}
    \caption{
      Multi-resolution reduction using medial axes.
      Top left uses level set $r=10$ and progresses from left to right across rows until $r=110$ at bottom right.
      Black shows input graph. 
      Grayscale background shows distance transform $D_G(x)$. 
      Thin white and thick white respectively show level sets $K_r$ and medial axes $M_r$.
    }
    \label{fig:medialAxis}
\end{figure*}

We computed various approximations of $M_r$ using the MATLAB functions \texttt{bwmorph} and \texttt{bwskel}. An example is shown in Fig.~\ref{fig:medialAxis}. 
Some of the results were compelling. For example, the graph in the top-right panel gives a convincing simplification of the original. 
Problems start to arise before long, though. 
The subsets of $K_r$ around C-shaped curves in the graph eventually join across the gap and produce a medial axis with unwanted cycles, as in the passage from the top-right panel to the middle-left panel. 
The reverse topological change of removing cycles can also cause issues. As the circular subset of $K_r$ within an existing cycle shrinks to a point and disappears, the medial axis collapses around it and suddenly jumps to pass through the center of the hole. This behavior is desirable for small-scale cycles, passing a single curve through the middle of a complex region as happens multiple times in the top row of the figure. When this happens in a large-scale cycle though, as in the two bottom-right panels, the medial axis departs entirely from the graph. 

A few other problems arose which were less dramatic but still need consideration. In many cases, we found the medial axis edges to be significantly more angular and linear than the organic originals. Also, imperfections in the smoothness of the level sets frequently caused terminal edges to split at their ends and stretch beyond the original graph out to the level set boundary (this issue is not present in the figure). 

In sum, this approach seems to hold some promise for moderate complexity reduction. The more serious topological issues don't arise until $r$ becomes large, and smaller $r$ values did produce some satisfying results. However, the value of $r$ where things break down is specific to each individual graph and can vary widely. Defining criteria to automatically determine an appropriate level of reduction could be difficult.

\section{Image Classification}
\label{sec:classification}

When using retinal images to detect disease, how much of the relevant information is carried by the fine details of the vessel networks, and how much can be found in the coarse large-scale structure alone? What is the relative importance of these two scales in recognizing retinal disease or distinguishing between neural cell types?
We investigated these questions with classification experiments.
To do this we reduced labeled graph datasets to a number of progressively lower resolutions, collected a set of graph statistics at each resolution level, and compared the accuracy of classification tasks performed using the statistics collected at different resolutions.
Informally, this might be thought of as a search for a border between signal and noise in the granularity of detail for a given data set. 
Steady classification rates in the presence of decreasing resolution would be interpreted to mean that details at that level of complexity are essentially noise, whereas sudden drops in classification rates would indicate that signal has been removed.

\subsection{Features Used for Classification}
\label{ssec:classificationFeatures}

\begin{figure*}
    \centering
    \includegraphics[width=1.1\textwidth, scale=1]{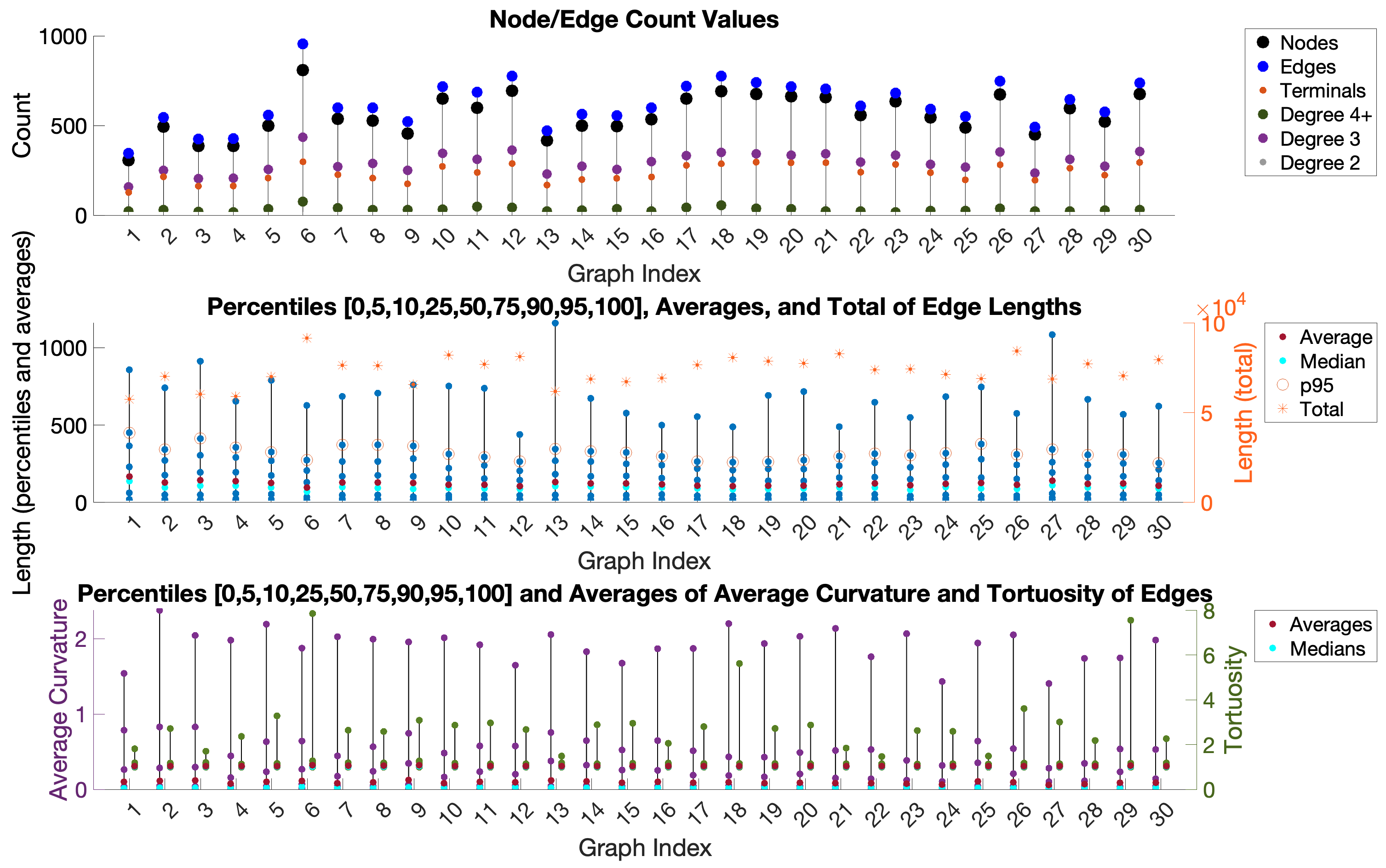}
    \caption{
      Graph statistics used as features for classification. 
      Each column (extended across rows) corresponds to a single graph.
      Statistics for 30 of the 45 unreduced graphs from the HRF data set are shown.
      Images 1-10 were labeled as healthy in the data set, images 11-20 were labeled as having diabetic retinopathy, and images 21-30 were labeled as glaucomatous.
    }
    \label{fig:graphStatistics}
\end{figure*}

The classification was performed using easily-interpretable statistics of the nodes and edges of each graph. The node-based statistics are counts of nodes of various degrees and the overall total. The edge-based statistics used percentiles, averages, and totals of edge length, average curvature, and tortuosity computed for each edge.
Table \ref{tbl:symbolsEqs} defines values used for computing the edge-based statistics, and Table \ref{tbl:featuresUsed} lists all of the statistics considered for each graph. 
Fig.~\ref{fig:graphStatistics} gives an example of statistics collected for 30 individual RBV graphs.
We performed our classification experiments twice: first using all 37 statistics listed in table \ref{tbl:featuresUsed}, and again separately using only the first 17, excluding average curvature and tortuosity. 

\begin{table*}
\centering
  \begin{tabular}{|c|l|} 
    \hline
    Edge in a graph 
      & $\beta:[0,1]\rightarrow\mathbb{R}^d$
    \\ \hline
    Length of an edge
      & $\ell = \int_0^1 \Vert\dot{\beta}(t)\Vert dt$
    \\ \hline
    Curvature at a point along an edge
      & $\kappa = \big\Vert\frac{dT}{ds}\big\Vert$
      \hspace{1mm} (unit tangent vector $T$ and arc length $s$)
    \\ \hline
    Average curvature of an edge
      & $\frac{1}{\ell} \int_0^{\ell}\kappa(s)ds$
    \\ \hline
    Tortuosity of an edge
      & $\tau = \ell \big/ \Vert \beta(1)-\beta(0) \Vert$
    \\ \hline
  \end{tabular}
  \caption{
      Values used to compute edge statistics
  }
  \label{tbl:symbolsEqs}
\end{table*}

\begin{table*}
\centering
  \begin{tabular}{|c|l|} 
    \hline
    1 
      & Total number of nodes  
    \\ \hline
    2-5
      & Nodes of degrees 1, 2, 3, and 4+
    \\ \hline 
    6
      & Number of edges 
    \\ \hline
    7
      & Sum of all edge lengths 
    \\ \hline
    8-16
      & Percentiles [0,5,10,25,50,75,90,95,100] of edge lengths
    \\ \hline
    17
      & Average edge length
    \\ \hline
    18-26
      & Percentiles [0,5,10,25,50,75,90,95,100] of average curvatures
    \\ \hline
    27
      & Average average curvature
    \\ \hline
    28-36
      & Percentiles [0,5,10,25,50,75,90,95,100] of tortuosities
    \\ \hline
    37
      & Average tortuosity
    \\ \hline
  \end{tabular}
  \caption{
      Features of each graph used for classification
  }
  \label{tbl:featuresUsed}
\end{table*}

\subsection{Data and Classification Methods}
\label{ssec:dataClass}

For the classification experiments using retinal blood vessels, we separately assessed small data sets from two sources: the Automated Retinal Image Analysis (ARIA) Data Set \cite{ariaData,FarnellBldVsl08} and the High-Resolution Fundus (HRF) Image Database \cite{hrfData,BudaiVesSeg13}. 
These data sets include fundus images that were converted to binary segmentations prior to our download, which we converted to shape graph format as described in section \ref{sec:dataProcessing}. 
From the ARIA data set, we used 57 images from subjects labeled as healthy, 17 labeled as  diabetic, and 18 labeled as having age-related macular degeneration (92 total). From the HRF data set, we used 15 images from subjects labeled as healthy, 15 labeled as having diabetic retinopathy, and 15 labeled as glaucomatous (45 total). Each image was given exactly one of these labels. 
Due to the limited sample sizes, we collapsed the disease categories and used the binary labels `healthy' and `disease' for our classification tasks.

For the classification experiments using neurons, we separately assessed two larger datasets from NeuroMorpho.org \cite{NeuroMorpho1,NeuroMorpho2} consisting of various types of rat neurons: the Althammer archive \cite{AlthammerMicrogliaAstrocytesRats20} and the Markram archive \cite{MarkramNeuropeptideRat05}. The NeuroMorpho archives contain data in \texttt{.swc} file format which describes the data in terms of three-dimensional locations, and it is more straightforward to convert these into shape graphs. For the neuron datasets, the goal was to classify the cell type according to provided labels. For the Althammer data set, we performed binary classification using 685 microglia cells from the hypothalamus and 181 oxytocinergic neurons from the hypothalamus (866 total). From the Markram data set, we performed multi-class classification using 239 basket cells from the neocortex, 35 double bouquet cells from the neocortex, 83 Martinotti cells from the neocortex, 60 neurogliaform cells from the neocortex, 427 pyramidal neurons from the neocortex, and 47 thalamocortical neurons from the thalamus (6 classes, 891 total).

For each RBV and neuron data set, we performed multi-resolution reduction as described in section \ref{sec:multiRes} to produce five sets of graphs: unreduced graphs with no changes beyond preprocessing, graphs with preliminary terminal trimming but no further reduction, and graphs at resolutions 0.8, 0.6, and 0.4. 
Statistics of the graphs were collected at each level of complexity. 
We then performed classification using support vector machines (SVM) from the Scikit-learn \cite{scikitLearn} package in Python 3 \cite{pythonRef09}. 
We used SVMs with a Gaussian radial basis function kernel and tried a grid of values for parameters $h$ and $\eta$ 
which respectively control the penalty on the slack variables and the variance in the Gaussian kernel. 
We used 18 values of $h$ including the numbers 0.5, 0.7, 0.9, and evenly-spaced values ranging from 1 to 29. 
For $\eta$ we used 250 evenly-spaced values ranging from 0.0001 to 0.1001. 
This parameter region was selected based on coarser preliminary computations as producing the highest-accuracy results. 
The classification accuracy for the RBV data sets was determined using leave-one-out cross-validation and the accuracy for the neuron data sets used ten-fold cross-validation. 
We performed the experiments using the full set of 37 features as well as using only the 17 features excluding average curvature and tortuosity.
For each graph set defined by data set, reduction level, and feature count, we computed the average, maximum, and standard deviation over all values in the grid of $h$ and $\eta$.

\subsection{Classification Results}
\label{ssec:classificationResults}

Breakdowns of the datasets and the results from the classification experiments are presented in Tables \ref{tbl:accuracy_rbv} and \ref{tbl:accuracy_neuron} for the RBV and neuron data respectively.
We emphasize that the primary goals here were to explore the use of interpretable geometry-based features and to analyze the relative importance of complex details versus general form in a shape graph representation. The aim was not to develop a state-of-the-art high-performance classifier.
In particular, these results are not comparable to pixel-based classifiers.
For reference, some recent approaches training neural networks on large data sets of color retinal fundus images have produced disease-detection classifiers that rival human experts, with accuracy rates in the high 90\% range for multiple retinal disorders in some cases \cite{DongRetinalOpticDiseaseScreening22,SonDeepLearnRetinalFundus19,JiangFundusDiseaseDNN,LeeAIDiabeticRetinopathy21}.
Concerning neurons, one recent paper tried a number of different deep neural network architectures to perform multi-class cell type classification on 35,000 rat neurons from NeuroMorpho.org and achieved accuracies around 90\% \cite{ZhangNeuronClassificationRat21}.

The classification performance for the RBV data was better than chance in all cases and substantially better in most, with maximum accuracy for selected parameter values at or above .8 and reaching as high as .933. 
Accuracy rates for the unreduced graphs were reliably among the highest, only ever being surpassed by the trimmed-only results. Accuracy for the trimmed-only graphs was usually higher than for the graphs with reduced resolution. 
However, neither of these was universally the case. For the HRF data using 17 features, the trimmed-only graphs had much better results than any other resolution level. We note that the average and maximum here of .829 and .933 were the highest for any condition, and came from a setup using only 17 of the 37 features. 
For the ARIA data with 37 features, the graphs with resolution .6 performed nearly as well as the unreduced graphs and outperformed all other reduced resolutions, while the trimmed-only graphs had the lowest maximum and second-lowest average.
There was no clear monotonic decrease in accuracy with resolution among the reduced-resolution RBV sets. The level .4 graphs had the lowest results among them in most cases, but these were usually not far from the next-lowest and included the highest average of the three for the HRF with 37 features.. The resolution .6 graphs had higher maximum accuracy than the resolution .8 graphs as often as not.

With the caveats that the results are not unanimous and that the data sets used are small, some general results are suggested. 
First, there is apparent potential in using these kinds of interpretable shape-based features to help detect retinal disease.
Second, much of the information relevant for performing classification based on the disease labels in these data sets lies in the terminal edges.
Third, reducing the complexity of the graphs using shapes that are similar-looking but not identical to the originals quickly removes most of the remaining information. That is, moderately reduced skeletons and heavily reduced skeletons are equally poor in providing diagnostic information, despite the moderate skeletons being visually much more similar to the original graphs than are the heavily reduced ones.

There were some qualitative differences in the results for the neuron data.
In the Althammer data set, the oxytocinergic neurons tended to have much denser graphs than the microglia and thus we considered the binary classification task to be an easy one. 
This was borne out in the results, with accuracy values ranging from mid to upper 90\% values. There was monotonic decrease in accuracy values without exception as the graph resolution was reduced. However, the accuracy loss was not very severe, and even the $.4$ resolution set frequently produced accuracy values above 95\%. 
Including the tortuosity and average curvature features improved the results everywhere except for the unreduced graphs, where the difference was very slight.

The multi-class classification in the Markram data set was a more difficult task. The classification rates were predictably lower than for the Althammer data, but were still much higher than chance. The classification rates gradually decreased from the trimmed-only graphs to the $.4$ resolution sets, but the lowest accuracies were not that much lower than the highest. Counterintuitively, the unreduced graphs had lower accuracies than the trimmed and $.8$ resolution sets.
Only minor differences were seen between the results using 17 versus 37 features.

In both of the neuron datasets, the results suggest that gross shape is more relevant than fine details in distinguishing between these cell types.


\begin{table*}
  \begin{subtable}[c]{0.45\textwidth}
  \centering
  \begin{tabular}{|c|c|}
    \hline
    \multicolumn{2}{|c|}{ARIA Data Classes}
    \\ \hline \hline
    Classification Label & Sample Size
    \\ \hline \hline
    Healthy & 57
    \\ \hline
    Disease & 35
    \\ \hline \hline
    Total & 92
    \\ \hline
  \end{tabular}
  \end{subtable}
  \begin{subtable}[c]{0.45\textwidth}
  \centering
  \begin{tabular}{|c|c|}
    \hline
    \multicolumn{2}{|c|}{HRF Data Classes}
    \\ \hline \hline
    Classification Label & Sample Size
    \\ \hline \hline
    Healthy & 15
    \\ \hline
    Disease & 30
    \\ \hline \hline
    Total & 45
    \\ \hline
  \end{tabular}
  \end{subtable}
  \\ \vspace{5mm}
  \begin{subtable}[c]{0.45\textwidth}
  \centering
  \begin{tabular}{|c|c|c|c|} 
    \hline
    \multicolumn{4}{|c|}{ARIA Accuracy, 37 Features}
    \\ \hline \hline
    & Average & Max & St.Dev
    \\ \hline
    Unreduced & 
      .689 & .783 & .030
    \\ \hline
    Trimmed only &
      .631 & .696 & .026
    \\ \hline
    $\rho=.8$ &
      .633 & .728 & .039
    \\ \hline
    $\rho=.6$ &
      .670 & .783 & .039
    \\ \hline
    $\rho=.4$ &
      .605 & .707 & .031
    \\ \hline
  \end{tabular}
  \end{subtable}
  \begin{subtable}[c]{0.45\textwidth}
  \centering
  \begin{tabular}{|c|c|c|c|} 
    \hline
    \multicolumn{4}{|c|}{HRF Accuracy, 37 Features}
    \\ \hline \hline
    & Average & Max & St.Dev
    \\ \hline
    Unreduced & 
      .776 & .889 & .059
    \\ \hline
    Trimmed only &
      .753 & .867 & .038
    \\ \hline
    $\rho=.8$ &
      .677 & .822 & .036
    \\ \hline
    $\rho=.6$ &
      .650 & .778 & .030
    \\ \hline
    $\rho=.4$ &
      .684 & .733 & .030
    \\ \hline
  \end{tabular}
  \end{subtable}
  \\ \vspace{5mm}
  \begin{subtable}[c]{0.45\textwidth}
  \centering
  \begin{tabular}{|c|c|c|c|} 
    \hline
    \multicolumn{4}{|c|}{ARIA Accuracy, 17 Features}
    \\ \hline \hline
    & Average & Max & St.Dev
    \\ \hline
    Unreduced & 
      .736 & .804 & .035
    \\ \hline
    Trimmed only &
      .590 & .652 & .026
    \\ \hline
    $\rho=.8$ &
      .543 & .652 & .037
    \\ \hline
    $\rho=.6$ &
      .575 & .663 & .026
    \\ \hline
    $\rho=.4$ &
      .537 & .641 & .032
    \\ \hline
  \end{tabular}
  \end{subtable}
  \begin{subtable}[c]{0.45\textwidth}
  \centering
  \begin{tabular}{|c|c|c|c|} 
    \hline
    \multicolumn{4}{|c|}{HRF Accuracy, 17 Features}
    \\ \hline \hline
    & Average & Max & St.Dev
    \\ \hline
    Unreduced & 
      .752 & .822 & .039
    \\ \hline
    Trimmed only &
      .829 & .933 & .041
    \\ \hline
    $\rho=.8$ &
      .645 & .800 & .066
    \\ \hline
    $\rho=.6$ &
      .681 & .756 & .041
    \\ \hline
    $\rho=.4$ &
      .625 & .689 & .036
    \\ \hline
  \end{tabular}
  \end{subtable}
  \\ \vspace{2mm}
  \caption{
    Data descriptions and classification results for RBV data sets.
    Accuracies were determined using leave-one-out cross-validation.
    Average, max, and standard deviation refer to repeated assessments with varying SVM parameters for slack penalty and Gaussian variance.
  }
  \label{tbl:accuracy_rbv}
\end{table*}


\begin{table*}
  \begin{subtable}[c]{0.45\textwidth}
  \centering
  \begin{tabular}{|c|c|}
    \hline
    \multicolumn{2}{|c|}{Althammer Data Classes}
    \\ \hline \hline
    Cell Type & Sample Size
    \\ \hline \hline
    Microglia & 685
    \\ \hline
    Oxytocinergic & 181 
    \\ \hline \hline
    Total & 866
    \\ \hline
  \end{tabular}
  \end{subtable}
  \begin{subtable}[c]{0.45\textwidth}
  \centering
  \begin{tabular}{|c|c|}
    \hline
    \multicolumn{2}{|c|}{Markram Data Classes}
    \\ \hline \hline
    Cell Type & Sample Size
    \\ \hline \hline
    Basket & 239
    \\ \hline
    Double Bouquet & 35
    \\ \hline
    Martinotti & 83
    \\ \hline
    Neurogliaform & 60
    \\ \hline
    Pyramidal & 427
    \\ \hline
    Thalamocortical & 47
    \\ \hline \hline
    Total & 891
    \\ \hline
  \end{tabular}
  \end{subtable}
  \\ \vspace{5mm}
  \begin{subtable}[c]{0.45\textwidth}
  \centering
  \begin{tabular}{|c|c|c|c|} 
    \hline
    \multicolumn{4}{|c|}{Althammer Accuracy, 37 Features}
    \\ \hline \hline
    & Average & Max & St.Dev
    \\ \hline
    Unreduced & 
      .996 & .999 & .004
    \\ \hline
    Trimmed only &
      .992 & .997 & .005
    \\ \hline
    $\rho=.8$ &
      .985 & .993 & .005
    \\ \hline
    $\rho=.6$ &
      .976 & .982 & .005
    \\ \hline
    $\rho=.4$ &
      .962 & .975 & .007
    \\ \hline
  \end{tabular}
  \end{subtable}
  \begin{subtable}[c]{0.45\textwidth}
  \centering
  \begin{tabular}{|c|c|c|c|} 
    \hline
    \multicolumn{4}{|c|}{Markram Accuracy, 37 Features}
    \\ \hline \hline
    & Average & Max & St.Dev
    \\ \hline
    Unreduced & 
      .667 & .708 & .019
    \\ \hline
    Trimmed only &
      .688 & .719 & .016
    \\ \hline
    $\rho=.8$ &
      .683 & .717 & .017
    \\ \hline
    $\rho=.6$ &
      .663 & .707 & .018
    \\ \hline
    $\rho=.4$ &
      .649 & .690 & .026
    \\ \hline
  \end{tabular}
  \end{subtable}
  \\ \vspace{5mm}
  \begin{subtable}[c]{0.45\textwidth}
  \centering
  \begin{tabular}{|c|c|c|c|} 
    \hline
    \multicolumn{4}{|c|}{Althammer Accuracy, 17 Features}
    \\ \hline \hline
    & Average & Max & St.Dev
    \\ \hline
    Unreduced & 
      .998 & 1.000 & .006 
    \\ \hline
    Trimmed only &
      .982 & .985 & .007
    \\ \hline
    $\rho=.8$ &
      .965 & .972 & .007
    \\ \hline
    $\rho=.6$ &
      .957 & .967 & .008
    \\ \hline
    $\rho=.4$ &
      .944 & .955 & .008
    \\ \hline
  \end{tabular}
  \end{subtable}
  \begin{subtable}[c]{0.45\textwidth}
  \centering
  \begin{tabular}{|c|c|c|c|} 
    \hline
    \multicolumn{4}{|c|}{Markram Accuracy, 17 Features}
    \\ \hline \hline
    & Average & Max & St.Dev
    \\ \hline
    Unreduced & 
      .678 & .699 & .015
    \\ \hline
    Trimmed only &
      .690 & .712 & .015
    \\ \hline
    $\rho=.8$ &
      .682 & .701 & .015
    \\ \hline
    $\rho=.6$ &
      .667 & .685 & .012
    \\ \hline
    $\rho=.4$ &
      .660 & .681 & .016
    \\ \hline
  \end{tabular}
  \end{subtable}
  \\ \vspace{2mm}
  \caption{
    Data descriptions and classification results for neuron data sets.
    Accuracies were determined using ten-fold cross-validation.
    Average, max, and standard deviation refer to repeated assessments with varying SVM parameters for slack penalty and Gaussian variance.
  }
  \label{tbl:accuracy_neuron}
\end{table*}

\section{Conclusions} 
\label{sec:conclusion}

This paper developed a method for reducing the detailed complexity of shape graphs while maintaining their fundamental shape. We demonstrated the method using shape graphs produced from real-world data in both two and three dimensions. We explored a variety of other reduction methods, a number of which apparently hold promise, and found the method spotlighted here to produce results that are visually more true to the originals.

The method was applied to disease detection in retinal fundus images and cell type classification of neurons by comparing classification accuracies using graphs at different degrees of complexity reduction, basing the classifications on interpretable geometric features. 
For both types of data, we found that these features did contain information relevant for classification. For the retinal blood vessels, a great deal of that information was associated with the terminal edges of the graphs and classification accuracy dropped quickly when complexity was reduced. For the neural cells, classification rates were robust to complexity reduction even with a great deal of simplification.

In our view, the method's subjective goal of extracting the basic form of a shape graph was convincingly achieved. It could be a useful tool in various applications where graph simplification is of interest.

\section*{Acknowledgements}

The authors acknowledge and thank the producers of the Althammer, ARIA, HRF, Markram, and Narkilahti datasets for making them public. This research was supported in part by the NSF grants NSF CDS\&E DMS 1953087 and NSF IIS 1955154.

\backmatter

\printbibliography
%
%

\end{document}